# Emergence of Superconductivity at 20 K in $Th_3P_4$-type $In_{3-x}S_4$ Synthesized by Diamond Anvil Cell with Boron-doped Diamond Electrodes

*R. Matsumoto[1], K. Yamane[1,2], T. Tadano[3], K. Terashima[1], T. Shinmei[4], T. Irifune[4], Y. Takano[1,2]

*Corresponding author; Email: MATSUMOTO.Ryo@nims.go.jp

[1]International Center for Materials Nanoarchitectonics (MANA),
National Institute for Materials Science, Tsukuba, Ibaraki 305-0047, Japan

[2]Graduate School of Pure and Applied Sciences, University of Tsukuba, 1-1-1 Tennodai, Tsukuba, Ibaraki 305-8577, Japan

[3]Research Center for Magnetic and Spintronic Materials,
National Institute for Materials Science, Tsukuba, Ibaraki 305-0047, Japan

[4]Geodynamics Research Center (GRC), Ehime University, Matsuyama, Ehime 790-8577, Japan

Keywords: Superconductivity, high pressure, Indium sulfide, diamond anvil cell

**Abstract**

The exploration of superconductors in metastable phases by manipulating crystal structures through high-pressure techniques has attracted significant interest in materials science to achieve a high critical temperature ($T_c$). In this study, we report an emergence of novel superconductivity in a metastable phase of $Th_3P_4$-type cubic $In_{3-x}S_4$ with remarkably high $T_c$ at 20 K under 45 GPa by using an originally designed diamond anvil cell equipped with boron-doped diamond electrodes, which can perform a high-pressure synthesis and an in-situ electrical transport measurement simultaneously. In-situ structural analysis indicates that the $In_{3-x}S_4$ appears partially above 40 GPa without heating. The high-pressure annealing treatment induces complete transformation to the $Th_3P_4$-type structure, and the defected concentration of $x$ in $In_{3-x}S_4$ decreases with increasing annealing temperature. The $T_c$ in $In_{3-x}S_4$ is maximized at $x = 0$ and approaches 20 K. Electronic band calculations show that the high density of states composed of sulfur and indium bands are located at the conduction band bottom near Fermi energy. The record high $T_c$ in $In_{3-x}S_4$ among superconducting sulfides accelerates the further exploration of high $T_c$ materials within the $Th_3P_4$-type cubic family by using flexibility in crystal structure.

# 1. Introduction

Since the discovery of superconductivity in mercury[1], the exploration of superconducting materials with higher transition temperature ($T_c$) has been continued, driven by their potential for ultimate energy-saving applications based on unique properties, such as zero resistivity. The $T_c$ has grown typically through a search for stable phases by varying the composed elements and synthesis temperature, as seen in the discovery of high-$T_c$ cuprates[2,3] and Fe-based materials[4]. However, such exploration has gradually tended to saturate due to the limited space of materials design. High-pressure application is a promising tool for expanding this space by stabilizing metastable phases by manipulating crystal structures. Most record $T_c$ values are achieved under high pressure in various superconducting families, such as $HgBa_2Ca_2Cu_3O_{8+\delta}$ ($T_c$ = 164 K at 20 GPa) in cuprates[5,6], La(O,F)BiS$_2$ ($T_c$ = 10 K at 1 GPa) in BiS$_2$-based materials[7,8], TaS$_2$ ($T_c$ = 16.4 K at 157.4 GPa) in transition-metal dichalcogenides[9], and LaH$_{10}$ ($T_c$ = 260 K at 188 GPa) in hydrides[10,11]. Recently discovered high-$T_c$s in nickelates of La$_3$Ni$_2$O$_7$[12] and La$_4$Ni$_3$O$_{10}$[13] also represent metastable phases under high pressure. Investigating novel metastable phases in different material groups has attracted considerable interest in discovering high-$T_c$ superconductors.

A group of compounds crystallizing in the Th$_3$P$_4$-type cubic structure offers significant freedom in designing functionality, as the Th and P sites can be replaced by various (cation)$^{4+}$ and (anion)$^{3-}$ species, such as Sr, Ti, Zr, Hf, Th, U, Np in the Th site, and N, P, As in the P site. Additionally, (cation)$^{3+}$ and (anion)$^{2-}$ can be accepted by introducing defects in the Th site, such as Y, lanthanoids (*Ln*), excluding Pm, Ac, Pu, Am, Cm, Bk, Cf in the Th site, and chalcogens (*Ch*) of S, Se, Te in the P site. By utilizing the compositional flexibility, various functional materials in this structure have been designed, including thermoelectric materials[14,15], ultrahard materials[16,17], and others[18]. The research for superconductivity in this structure has been conducted before the discovery of cuprate, and relatively high $T_c$ was reported. In particular, La$_{3-x}$*Ch*$_4$ shows superconductivity at ambient pressure with a $T_c$ of 8.3 K in La$_{3-x}$S$_4$[19] and 8.5 K in La$_{3-x}$Se$_4$[20]. Recent first-principles calculations predict the emergence of superconductivity in several metastable Th$_3$P$_4$-type materials under high pressure[21–23]. Among these predictions, Sn$_3$S$_4$ is estimated to have the highest $T_c$ [22], and the superconductivity with $T_c$ of 12 K at 5.6 GPa is confirmed experimentally through high-pressure synthesis.[24]. However, the exploration of superconductors in the metastable phase of this family with varying compositions is still an open issue due to the difficulty in the sample synthesis and in-situ measurements of physical properties under high pressure.

In this study, we focus on the Th$_3$P$_4$-type cubic In$_{3-x}$S$_4$ as a candidate for a superconductor. The stable phase at ambient pressure for the In$_{3-x}$S$_4$ is the layered tetragonal structure of In$_2$S$_3$[25]. The cubic In$_{3-x}$S$_4$ is obtained as a metastable phase via high-pressure annealing of the tetragonal In$_2$S$_3$[26]. The defect concentration of $x$ in In$_{3-x}$S$_4$ is typically 0.33 when the In$_2$S$_3$ is homogeneously transformed to the Th$_3$P$_4$-type structure. Even though a similar compound of Th$_3$P$_4$-type In$_{3-x}$Se$_4$ is obtained from In$_2$Se$_3$ by the compression and exhibit superconductivity[27], knowledge of the physical properties, particularly the emergence of superconductivity, is still lacking for In$_{3-x}$S$_4$. Understanding the physical properties of In$_{3-x}$S$_4$ provides valuable insights for future exploration of superconductors in this family. We have conducted high-pressure annealing and in-situ characterization for tetragonal In$_2$S$_3$ under various conditions and obtained Th$_3$P$_4$-type In$_{3-x}$S$_4$ using a diamond anvil cell (DAC) high-pressure apparatus. The crystal structure, including the defected amount of In, was quantitively investigated using synchrotron X-ray diffraction (SXRD) analysis in the DAC during compression. In-situ electrical transport measurements of the obtained In$_{3-x}$S$_4$ were performed in the same chamber of DAC used for SXRD analysis. Our high-pressure research successfully reveals that Th$_3$P$_4$-type In$_{3-x}$S$_4$ exhibits a surprisingly high $T_c$ of 20 K. The superconducting properties are strongly related to the deficiency in the In site. The first-principles calculations indicated that the electronic states of the non-defected In$_3$S$_4$ are mainly composed of In and S bands with comparable contribution, which totally differs from typically known high-$T_c$ sulfides. The discovery of superconductivity in In$_{3-x}$S$_4$ opens avenues for further exploration of advanced materials, as its $T_c$ is the highest record among binary sulfides, except for high-$T_c$ hydrides[28,29].

## 2. Results and discussion
### 2.1 Synthesis of Th$_3$P$_4$-type In$_{3-x}$S$_4$

Th$_3$P$_4$-type In$_{3-x}$S$_4$ was synthesized via high-pressure annealing for tetragonal In$_2$S$_3$ using a specially designed DAC, as shown in Fig. 1 (a). Boron-doped diamond (BDD) with a high boron concentration above $10^{21}$ cm$^{-3}$, which exhibits metallic transport property[30], is fabricated onto the surface of the diamond anvil as electrodes for electrical measurements[31–33]. The BDD heater and thermometer are positioned near the sample space for temperature control during high-pressure annealing[24,34]. A small piece of tetragonal In$_2$S$_3$ is placed in the center of the diamond anvil, as depicted in Fig. 1 (b). The sample was compressed by squeezing the DAC to above 30 GPa and then annealed using the BDD heater while measuring electrical resistance in the DAC. Figure 1 (c) shows a typical behavior of resistance and temperature sequence during high-pressure annealing to obtain Th$_3$P$_4$-type In$_{3-x}$S$_4$. The compressed In$_2$S$_3$ initially exhibits semiconducting behavior with decreasing resistance as temperature increases. At around 350 K, the resistance continues to reduce even with constant temperature, indicating progress of structural phase transition from tetragonal In$_2$S$_3$ to a lower resistance phase. Consequently, metallic properties with increasing resistance against temperature are observed above 400 K and the cooling process.

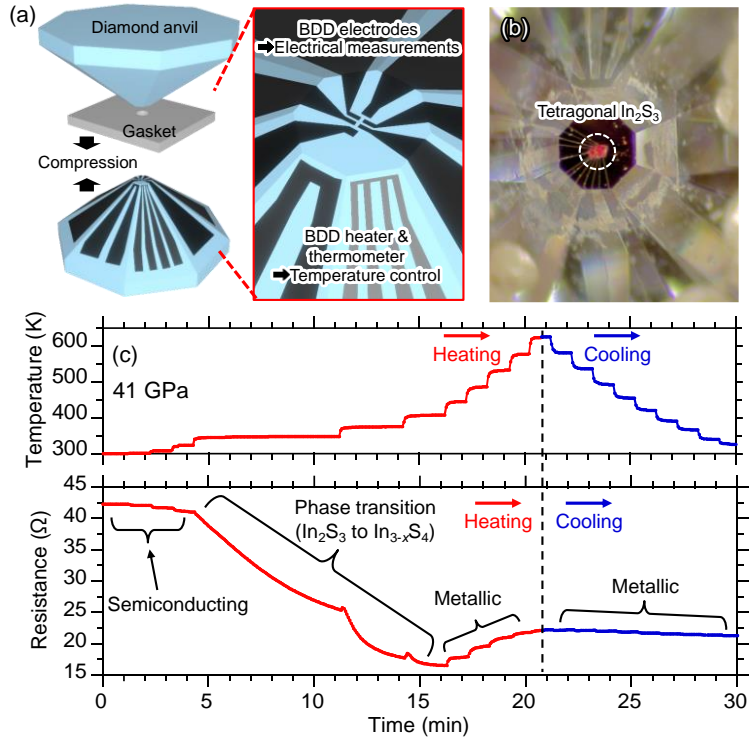

FIG. 1. (a) Schematic image of diamond anvil cell (DAC) with boron-doped diamond (BDD) components. (b) Optical microscope image of the mounted tetragonal $In_2S_3$ onto the diamond anvil. (c) Typical behavior of the resistance during the high-pressure annealing to obtain $Th_3P_4$-type $In_{3-x}S_4$. The upper figure is the temperature sequence, and the lower one is the change in the electrical resistance of the sample.

Figure 2 presents a typical SXRD pattern along with the result from Rietveld refinement for the sample at 45 GPa after high-pressure annealing. The green and gray bars indicate simulated peak positions for $Th_3P_4$-type $In_3S_4$ and cubic BN as a pressure-transmitting medium, respectively. The blue spectrum represents the differential curve in the fitting. The analysis reveals that the sample crystallizes with a $Th_3P_4$-type cubic structure ($I$-$43d$) with no detectable impurity phases. The derived lattice parameter is $a = 7.5309$ Å with a reliability factor $R_{wp} = 0.817\%$, which qualitatively agrees with a previous report of the high-pressure synthesis of $Th_3P_4$-type $In_{3-x}S_4$ via laser heating in a DAC[26]. Although these SXRD results are obtained from the sample sintered under 44 GPa and 1100 K, the $Th_3P_4$-type structure has appeared at lower pressure and temperature conditions below 41 GPa and 600 K, as shown in Fig. S1. These results suggest that the phase transition from tetragonal $In_2S_3$ to $Th_3P_4$-type $In_{3-x}S_4$ occurs near the boundary of semiconductor to metal transition, as discussed in Fig. 1 (c). Additionally, the sample partially exhibits a signature of a $Th_3P_4$-type diffraction pattern even before annealing, indicating that the $In_{3-x}S_4$ phase begins to form at room temperature. A similar trend of pressure-induced transformation to the $Th_3P_4$-type phase without heating has been observed for $In_{3-x}Se_4$[27]. The $Th_3P_4$-type structure rapidly decomposes with decreasing pressure, as shown in the SXRD patterns during the decompression process in Fig. S2. The behavior of instability in $In_{3-x}S_4$, indicating a low energy barrier between the low-pressure phase and Th$_3$P$_4$-type structure, contrasts significantly with related compounds such as Sn$_3$S$_4$, which requires annealing at above 30 GPa to achieve the Th$_3$P$_4$-type structure and remains stable down to 5 GPa[24].

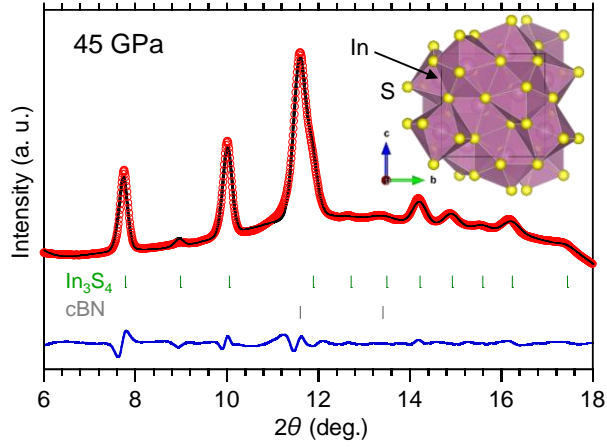

**FIG. 2. SXRD patterns with synchrotron radiation ($\lambda$ = 0.4171 Å) of obtained In$_{3-x}$S$_4$ under 45 GPa with the fitting result of Rietveld refinement. The green and gray bars indicate the peak positions of In$_3$S$_4$ and cubic BN as a pressure-transmitting medium. The blue spectrum means a differential curve for the fitting. The schematic image of the crystal structure of In$_{3-x}$S$_4$ is shown in the inset.**

## 2.2 Superconducting properties

Figure 3 (a) shows the temperature ($T$) dependence of resistance ($R$) up to 300 K in Th$_3$P$_4$-type In$_{3-x}$S$_4$ at 45 GPa. The same sample of the SXRD analysis, as indicated in Fig. 2, is used for the electrical measurement. An expanded plot of the low-temperature region is displayed in the inset of Fig. 3 (a). The Th$_3$P$_4$-type In$_{3-x}$S$_4$ exhibits metallic transport properties with a clear drop in resistance to zero, corresponding to superconductivity emerging around 20 K. As shown in the inset, the resistance curve in the low-temperature region is well-fitted with the Bloch-Gruneisen (BG) equation, described by the following equation[35],

$$R(T)=R_0+A\left(\frac{T}{\theta_\mathrm{D}}\right)^5 \int_0^{\theta_\mathrm{D}/T} \frac{x^5}{(e^x-1)(1-e^{-x})}dx$$

where $R_0$ is the residual resistance, $A$ is a characteristic constant, and $\theta_\mathrm{D}$ is the Debye temperature. The parameters are $R_0$ = 0.0333 Ω, $A$ = 0.141, and $\theta_\mathrm{D}$ = 171 K. The estimated $\theta_\mathrm{D}$, which serves as a proportionality constant for $T_\mathrm{c}$ within Bardeen-Cooper-Schrieffer (BCS) theory[36,37], is significantly lower than that of other superconducting families with comparable $T_\mathrm{c}$, such as A15-type compounds[38]. One possible reason for the high $T_\mathrm{c}$ in In$_{3-x}$S$_4$ is the high electronic density of state (DOS) at Fermi energy ($E_\mathrm{F}$), which generally enhances the $T_\mathrm{c}$[39]. Additionally, the sample before high-pressure annealing also exhibits a superconducting transition at 9 K, although the $R$-$T$ curve is non-metallic and zero resistance is not observed, as shown in Fig. S3. According to previous SXRD analyses, tetragonal $In_2S_3$, which is stable at ambient conditions, undergoes several structural phase transitions under high pressure from phase I to phase III to an amorphous state[25,26,40]. In our SXRD analysis, amorphous-like broadened diffraction peaks are consistently observed before annealing at 41 GPa. On the other hand, small peaks corresponding to the $Th_3P_4$-type structure are also visible in the diffraction patterns before annealing. These observations indicate that the signature of superconductivity with $T_c$ = 9 K in as-pressed sample also originates from $Th_3P_4$-type $In_{3-x}S_4$, and the $T_c$ significantly enhances to 20 K via annealing. The modulation of $T_c$ is possibly related to changes in the amount of In defects because similar modifications in defect amounts occur in other $Th_3P_4$-type compounds, affecting their physical properties, such as in superconducting $Y_{3-x}S_4$[41,42] and high-temperature thermoelectric material $La_{3-x}Te_4$[43].

Figure 3 (b) depicts the enlarged $R$-$T$ curve of $Th_3P_4$-type $In_{3-x}S_4$ in the low-temperature region under various magnetic fields to detail the superconducting transition. The inset shows a more enlarged plot around the starting temperature of the drop in resistance, defined as $T_c^{onset}$. Comparison of the $R$-$T$ curves under 0 and 7 T reveals that the $T_c^{onset}$ of $In_{3-x}S_4$ at 45 GPa is 20.0 K, while the temperature at zero resistance ($T_c^{zero}$) is 17.2 K. The superconducting transition is gradually suppressed by applying magnetic fields up to 7 T. The upper critical field $\mu_0H_{c2}(0)$ is estimated using the temperature dependence of $\mu_0H_{c2}$, as shown in Fig. 3 (c). The criterion used to determine $T_c$ under each magnetic field is the temperature at 99% of normal resistance. The plot with Werthamer-Helfand-Hohenberg (WHH) fitting[44,45] reveals a $\mu_0H_{c2}(0)$ of 31.7 T, which is comparable to the weak-coupling Pauli limit ($1.84T_c$ = 36.4 T). This behavior contrasts with isostructural $La_{3-x}S_4$, which exhibits higher $\mu_0H_{c2}(0)$ than $1.84T_c$[46]. The coherence length at zero temperature $\xi(0)$ is determined to be 3.2 nm from the Ginzburg-Landau (GL) formula $\mu_0H_{c2}(0) = \Phi_0/2\pi\xi(0)^2$, where the $\Phi_0$ is a fluxoid.

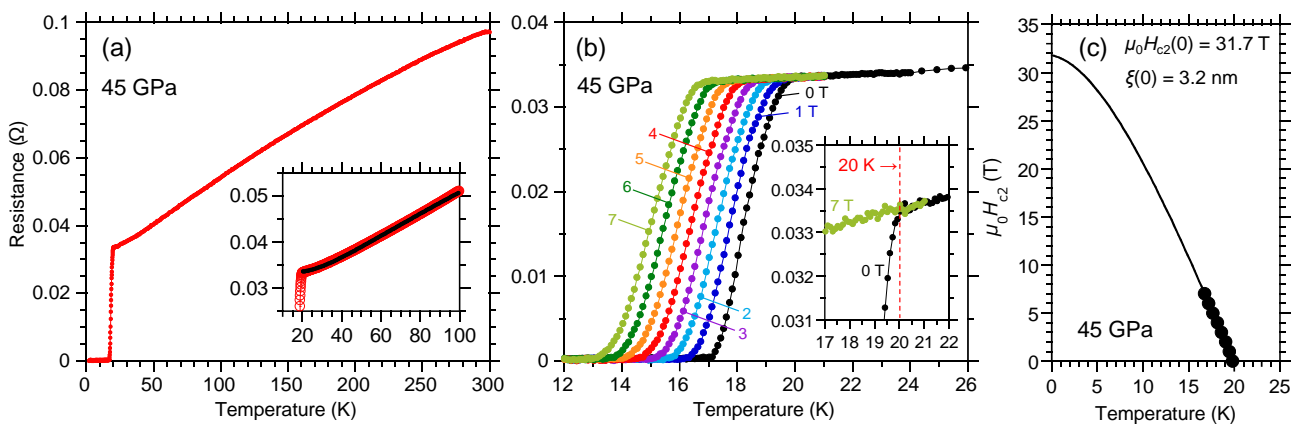

**FIG. 3. Results of electrical transport measurements under high-pressure in obtained $In_{3-x}S_4$. (a) Temperature ($T$) dependence of resistance ($R$) at 45 GPa. The inset shows the enlarged plot below 100 K with BG fitting. (b) $R$-$T$ curves under magnetic fields. The onset $T_c$ is defined by the separating point of the $R$-$T$ curve under 0 and 7 T, as presented in the inset. (c) Temperature dependence of $\mu_0H_{c2}$.**

Figure 4 (a) shows the variation of $T_c$ in $In_{3-x}S_4$ under different annealing conditions from our experimental runs. The remarkably high $T_c$ in $In_{3-x}S_4$ is reproducibly observed. Moreover, a positive trend between $T_c$ and annealing temperature is indicated across a range of 14 to 20 K, although pressure values in each plot are slightly different. To investigate the reasons for $T_c$ variation, we performed an SXRD analysis for each sample after annealing in Run5 and estimated the amount of In deficiency based on occupancy analysis using Rietveld refinements. Figure 4 (b) presents the relationship between the amount of In defect and $T_c$ in $In_{3-x}S_4$, with labeled values representing the annealing temperature. The amount of In defects $x$ decreases monotonically with the increase of annealing temperature and approaches zero in the sample sintered at 1100 K, which has the highest $T_c$ of 20 K. Additionally, the $T_c$ in as-pressed sample is plotted as $In_{2.67}S_4$, and the plot smoothly connects to the annealed samples. These observations suggest that the starting material $In_2S_3$ transforms into superconducting $Th_3P_4$-type $In_{3-x}S_4$ with $T_c$ of 9 K through compression without the annealing treatment. The $T_c$ of $Th_3P_4$-type $In_{3-x}S_4$ can be widely tuned up to 20 K by modifying the amount of In deficiency via high-pressure annealing. A potential concern is the generation of pure sulfur as an impurity in the sample chamber, known as a superconductor with $T_c$ of 15 K[47] because the initial $In_{2.67}S_4$ transforms to $In_3S_4$ during sintering. We believe that the observed high $T_c$ of 20 K originates from $In_3S_4$, as the metallic phase of sulfur-III with an orthorhombic structure should appear above 80 GPa, which is significantly higher than our experimental conditions[48–50].

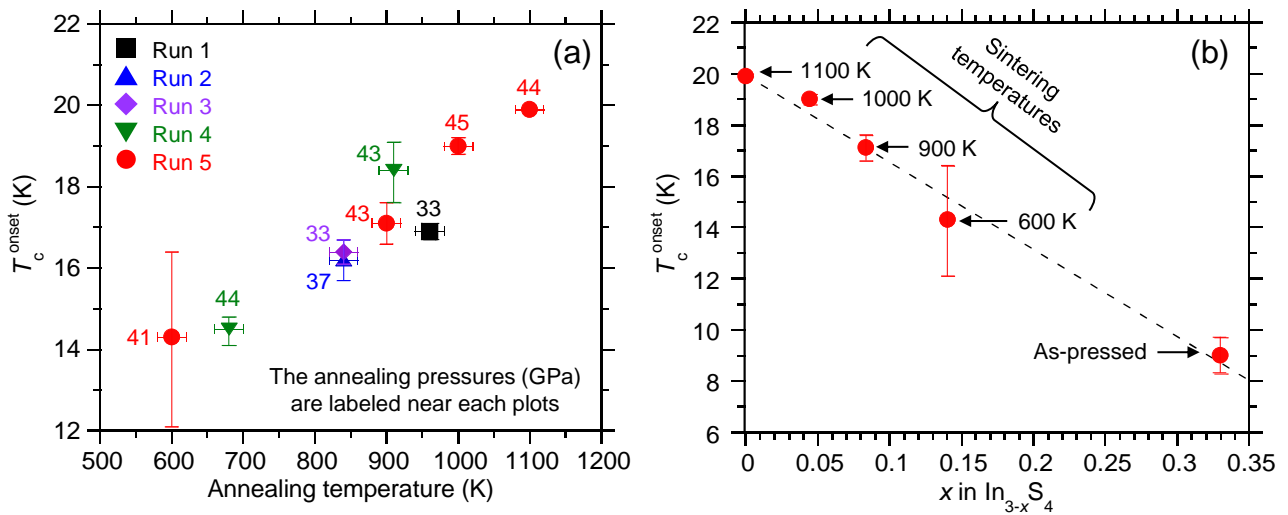

**FIG. 4. (a) The obtained $T_c$ in $In_{3-x}S_4$ sintered by various temperatures. The annealing pressures (GPa) are labeled near each plot. (b) In amount dependence of $T_c$ in $In_{3-x}S_4$. The dashed line is a guide for the eye.**

## 2.3 Calculation of electronic structure

Based on the estimated amount of In, the electronic band structure of $Th_3P_4$-type $In_3S_4$, without In deficiency, at 40 GPa is depicted in Fig. 5 (a). Several electron bands cross the $E_F$, indicating a metallic electronic state, which is consistent with our electrical transport measurements. The inset in Fig. 5 (b) presents the typical Hall resistance as a function of the applied magnetic field to determine the carrier type for the $In_{3-x}S_4$ obtained from transport measurements. The negative slope of the Hall resistance versus magnetic field indicates an n-type characteristic, which aligns with the results of band calculation. Figure 5 (b) shows the electronic DOS projected onto atomic orbitals. The bands crossing $E_F$ of $In_3S_4$ are mainly composed of In $s$, In $p$, and S $p$ orbitals. Notably, the comparable contributions of In $s$ and S $p$ orbitals provide a high total-DOS at $E_F$ of 10 states/eV/unit cell at the conduction band bottom, which is twice that of $Sn_3S_4$[24]. The sharp peak in DOS near $E_F$, resembling a van-Hove singularity (vHs), is advantageous for achieving high-$T_c$, similar to hydrides[39]. On the other hand, such singularity in the DOS possibly induces structural instability. The observed high $T_c$ and rapid decomposition against a pressure change in $In_{3-x}S_4$ may be attributed to the vHs-like electronic states. Also, the complex DOS with sharp peaks is consistent with the dramatic changes in $T_c$ observed in our experiments because the metal composition influences the carrier concentration, namely the position of $E_F$. Additionally, a notable insight in the electronic states is the existence of an extremely high DOS, exceeding 23 states/eV/unit cell, located at the valence band top. This band is composed of only S $p$ orbitals, similar to the electronic state of superconducting sulfur above 100 GPa. In a hole-doped $In_{3-x}S_4$, the realization of sulfur-dominant superconductivity at lower pressure is anticipated, such as in the case of high-$T_c$ hydrides through the tuning of $E_F$ position via the defect-engineering and elemental substitution due to the unique feature of high degree of freedom in composing elements of $Th_3P_4$-type family.

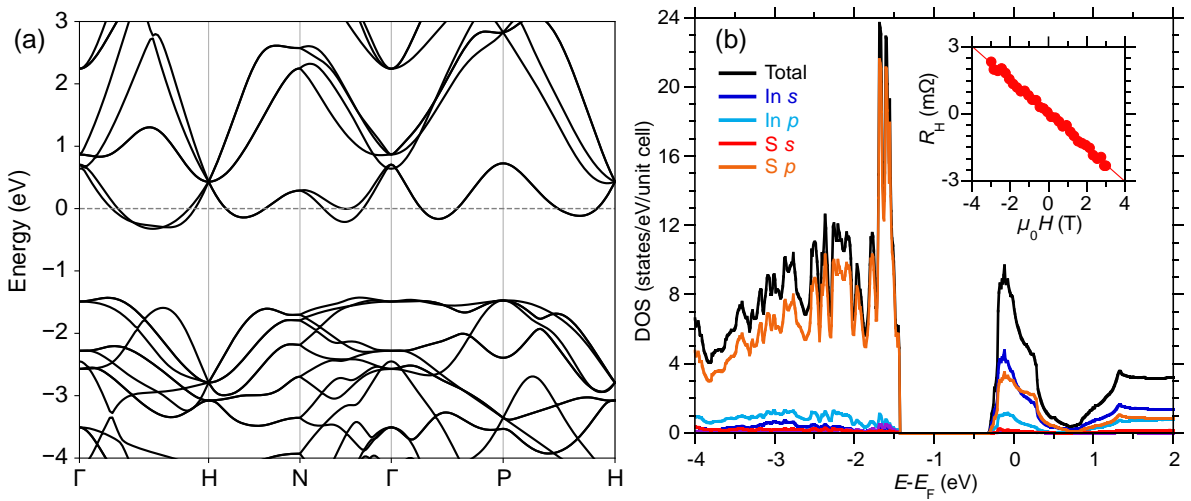

**FIG. 5. (a) Electronic band structure in $In_3S_4$ at 40 GPa. (b) Electronic density of states projected onto atomic orbitals in $In_3S_4$ at 40 GPa. The inset shows typical Hall resistance as a function of the applied magnetic field.**

## 2.4 $T_c$-$P$ diagram in various sulfides

Figure 6 displays the relationship between $T_c$ and applied pressure ($P$) in various superconducting binary sulfides[51–59], Th$_3$P$_4$-type superconductors[24,51,60], and pure sulfur[61], as referring by a summary from previous paper[51]. In$_3$S$_4$ obviously exhibits a higher $T_c$ than other isostructural Th$_3$P$_4$-type materials. It also holds the highest $T_c$ record among all discovered sulfide superconductors, except for the high-$T_c$ hydride H$_3$S[62]. PbS has recently been reported to show a high $T_c$ in sulfides at a moderate pressure of 19 GPa[51]. Electronic band calculations indicate that PbS has a similar electronic structure around $E_F$ as TaS$_2$ and pure sulfur under high pressure, both of which show high $T_c$ above 15 K and a significant contribution of the sulfur band to $E_F$[51]. Therefore, the critical factor for achieving high $T_c$ in sulfides is considered to be the sulfur band, which replicates the superconducting sulfur. Conversely, In$_3$S$_4$ exhibits a higher $T_c$ than these sulfides at lower pressure regions despite comparable contributions of S and In bands to $E_F$. Further investigation into the origin of the high $T_c$ and the role of In is expected as further research topics. Recently, C. J. Pickard et al. proposed a figure of merit $S$-value for an evaluation of pressure-induced superconductivity as $S=\sqrt{T_c/T_{c,MgB_2}^2 + P^2}$, where the $T_c$ of MgB$_2$ is 39 K, and the $S$-value should be 1 in the case of MgB$_2$ at ambient pressure[63]. Among binary sulfides, Th$_3$P$_4$-type In$_3$S$_4$ and Sn$_3$S$_4$ exhibit high $S$-value beyond 0.3, as shown in the dashed line of Fig. 6. This high $S$-value accelerates the exploration of practical superconducting materials within the metastable phases of the Th$_3$P$_4$-type cubic family.

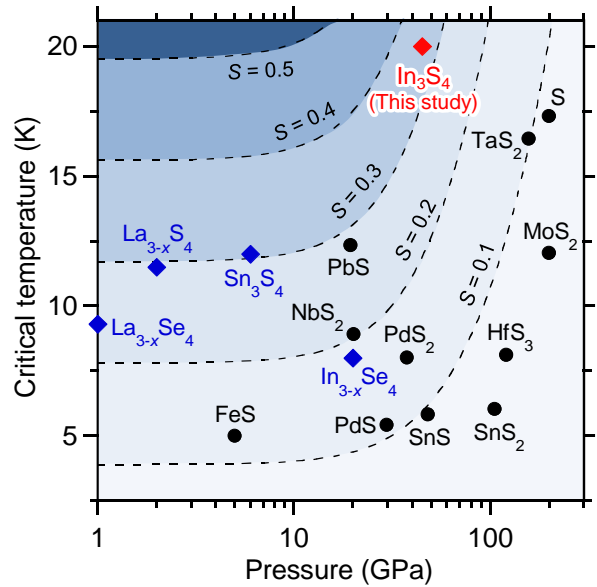

**FIG. 6. The $T_c$ diagram against applied pressure in various superconducting binary sulfides, Th$_3$P$_4$-type superconductors, and pure sulfur[51–61], as referring by a summary from previous paper[51]. The dashed lines indicate a border for a figure of merit $S$ for pressure-induced superconductivity.**

## 3. Conclusion

The highest $T_c$ among the superconducting materials family has typically been discovered under high pressure. Due to their high degree of freedom in composing elements and tunable functionality, we have explored $Th_3P_4$-type cubic materials as a new vein for the superconducting family. This study demonstrates a series of experiments involving high-pressure synthesis, in-situ structural analysis, and electrical transport measurements of $Th_3P_4$-type $In_{3-x}S_4$ using a custom-designed DAC with BDD electrodes and heater. Key achievements of this study include i) the emergence of superconductivity in $In_{3-x}S_4$, ii) the anomalous enhancement of $T_c$ up to 20 K, and iii) a unique electronic band structure.

First, the starting material of tetragonal $In_2S_3$ undergoes several pressure-induced structural phase transitions and partially transforms into $Th_3P_4$-type $In_{3-x}S_4$ above 40 GPa without external heating. Although pressure-driven $In_{3-x}S_4$, namely $In_{2.67}S_4$, exhibits superconductivity at 9 K, its $R$-$T$ curve remains semiconducting, and zero resistance is not observed. Thus, the transformation into superconducting $Th_3P_4$-type $In_{3-x}S_4$ is incomplete without annealing. Second, high-pressure annealing induces a clear metallic property, zero resistance in the superconducting state, and a single-phase SXRD pattern of $Th_3P_4$-type $In_{3-x}S_4$. The amount of In defect ($x$) in $In_{3-x}S_4$ systematically reduces with the increase of annealing temperature. A monotonic enhancement of $T_c$ is observed with an optimal highest $T_c$ of 20 K in $In_3S_4$ without In defects. This unexpectedly high $T_c$ is the highest record among binary superconducting sulfides, excluding $H_3S$. Finally, $In_3S_4$ exhibits a high DOS with comparable contributions from In and S orbitals at the conduction band bottom near $E_F$, which differs from the electronic states of other high $T_c$ sulfides. Additionally, a much higher DOS is located at the valence band top, composed of only the sulfur contribution.

Our findings on superconductivity in $In_{3-x}S_4$ open new avenues for both experimental and theoretical research fields to realize higher $T_c$ in metastable materials because the $Th_3P_4$-type structure has high tunability in composing elements. Further exploration of related compounds in the $Th_3P_4$-type family is explored to understand the mechanism of anomalous superconductivity in $In_{3-x}S_4$.

**Experimental section**

**Preparation of DAC:** For DAC experiments, the starting materials were placed into a sample chamber of DAC equipped with BDD electrodes[31–33] and heater[34] for high-pressure annealing, in-situ SXRD analysis, and electrical measurements. Single crystalline and nano-polycrystalline diamonds [64] were used for the anvil material. A Re sheet and cubic BN served as the sample chamber and pressure-transmitting medium. The applied pressure was estimated from the fluorescence of ruby powder[65] and peak shift in the Raman spectrum of the diamond anvil tip[66] using an inVia Raman Microscope (RENISHAW). $R$-$T$ measurements were performed in a physical property measurement system (PPMS, Quantum Design) with a 7 T superconducting magnet.

**High-pressure synthesis:** Reproducible high-pressure synthesis was conducted through runs 1 to 5. High-pressure synthesis was performed in all runs using a custom-designed DAC with BDD components. In the run1, a mixture of orthorhombic InS and tetragonal $In_2S_3$ were used as starting materials with a stoichiometric composition of In:S=3:4. InS and $In_2S_3$ were synthesized via conventional melt and slow-cooling methods using In and S in an evacuated quartz tube. To obtain $Th_3P_4$-type $In_{3-x}S_4$, high-pressure annealing was performed at 33 (35) GPa and 960 K. In the run 2 to 5, only $In_2S_3$ was used for the synthesis. Conditions were 37 (38) GPa and 840 K in the run2, 33 (25) GPa and 840 K in the run3. In the runs 4 and 5, annealing was conducted several times for one sample. In the run4, the first sintering conditions were 44 (51) GPa and 680 K, and the second was 43 (43) GPa and 910 K. In the run5, the first sintering at 41 (43) GPa and 600 K, the second at 43 (45) GPa and 900 K, the third at 45 (44) GPa and 1000 K, and the fourth at 44 (45) GPa and 1100 K were performed. Pressure values naturally varied during the annealing cycles and those after sintering were indicated inside the parentheses. After all annealing treatments, the temperature dependence of resistance in the obtained samples was measured. The in-situ SXRD analysis was conducted after the annealing in the runs 3 and 5.

**Structural analysis:** SXRD patterns and *R-T* curves were measured under corresponding pressures. SXRD measurements were carried out using synchrotron radiation at the AR-NE1A beamline in the Photon Factory (PF) located at the High Energy Accelerator Research Organization (KEK). The energy of X-ray beam was monochromatized to 30 keV ($\lambda = 0.4171$ Å). The X-ray is introduced to the sample in the DAC through a collimator with 50 μm diameter. SXRD patterns were integrated into a one-dimensional profile using IPAnalyzer, and lattice constants were determined using PDIndexer[67]. SXRD patterns were refined by Rietveld analysis using RIETAN-FP software[68] to estimate the occupancy in metal sites. Crystal structure images were generated using VESTA software[69].

**Theoretical calculation:** Electronic structures at high pressure was calculated using Quantum ESPRESSO (QE)[70–72]. The generalized gradient approximation (GGA) of Perdew–Burke–Ernzerhof (PBE)[73] was used to describe the exchange-correlation function with the pseudopotentials obtained from the SSSP PBE Efficiency v1.3.0 library[74]. Stable atomic positions and lattice constants were calculated under pressure before the electronic structure calculations. A $8 \times 8 \times 8$ *k*-grid was employed for the *k*-point sampling in the first Brillouin zone, and the kinetic energy cutoffs for the expansion of the electronic wave function were set to 50 Ry. A *k*-point mesh of $16 \times 16 \times 16$ was used to calculate the DOS.

**ACKNOWLEDGMENTS**

This work was partly supported by JSPS KAKENHI Grant Number 23H01835, 23K13549, and 23KK0088. The fabrication process of diamond electrodes was partially supported by the NIMS Nanofabrication Platform in the Nanotechnology Platform Project sponsored by the Ministry of Education, Culture, Sports, Science and Technology (MEXT), Japan. The nano-polycrystalline diamond was synthesized and provided via the Visiting Researcher's Program of the GRC with proposal No. 2023YB01. The synchrotron X-ray experiments were performed at AR-NE1A (KEK-PF) under the approval of proposal No. 2022G049 and 2024G084 with support from Dr. Y. Shibazaki (KEK). This work was supported by the World Premier International Research Center Initiative (WPI), MEXT, Japan.

**Table of contents**

High-pressure synthesis and in-situ measurement of crystal structure and electrical transport properties using custom-design diamond anvil cell with boron-doped diamond electrodes reveals that $Th_3P_4$-type $In_{3-x}S_4$ exhibits an emergence of superconductivity with high $T_c$ of 20 K under high pressure. The $T_c$ is the highest record among all the superconducting sulfides except for hydrides.

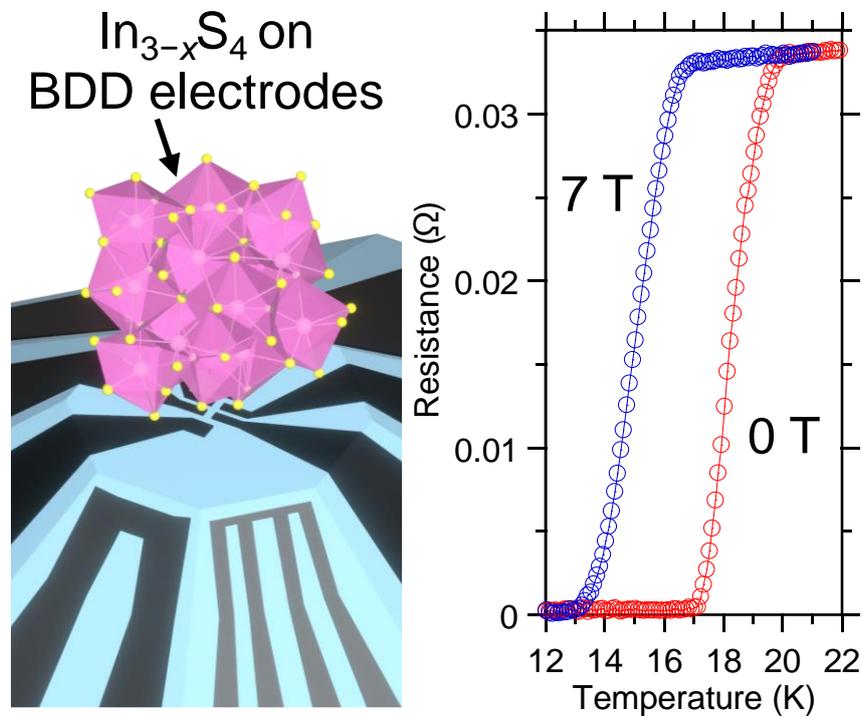

# Supplemental information for Emergence of Superconductivity at 20 K in Th$_3$P$_4$-type In$_{3-x}$S$_4$ Synthesized by Diamond Anvil Cell with Boron-doped Diamond Electrodes

*R. Matsumoto[1], K. Yamane[1,2], T. Tadano[3], K. Terashima[1], T. Shinmei[4], T. Irifune[4], Y. Takano[1,2]
*Corresponding author; Email: MATSUMOTO.Ryo@nims.go.jp

[1]International Center for Materials Nanoarchitectonics (MANA),
National Institute for Materials Science, Tsukuba, Ibaraki 305-0047, Japan
[2]Graduate School of Pure and Applied Sciences, University of Tsukuba, 1-1-1 Tennodai, Tsukuba,
Ibaraki 305-8577, Japan
[3]Research Center for Magnetic and Spintronic Materials,
National Institute for Materials Science, Tsukuba, Ibaraki 305-0047, Japan
[4]Geodynamics Research Center (GRC), Ehime University, Matsuyama, Ehime 790-8577, Japan

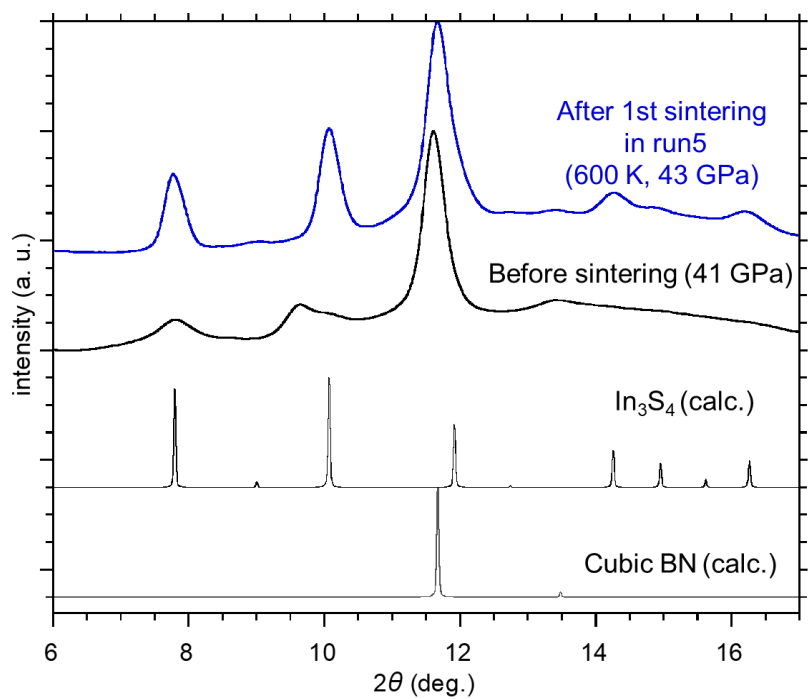

**FIG. S1. Comparison of SXRD patterns between as-pressed In$_2$S$_3$ and sintered one under 41 GPa. The calculation patterns of Th$_3$P$_4$-type In$_3$S$_4$ cubic BN are also shown.**

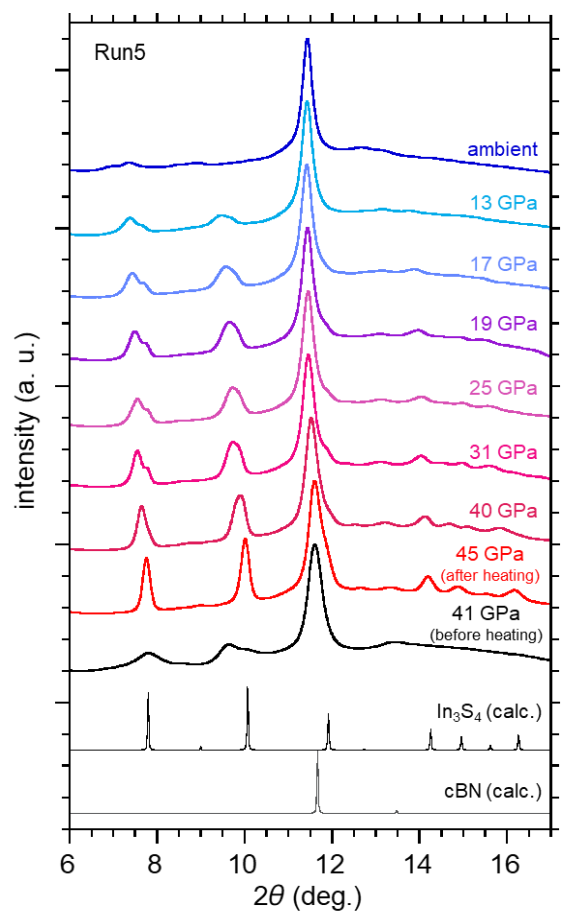

**FIG. S2.** SXRD patterns obtained $In_{3-x}S_4$ in the decompression process. The calculation patterns of $Th_3P_4$-type $In_3S_4$ cubic BN are also shown.

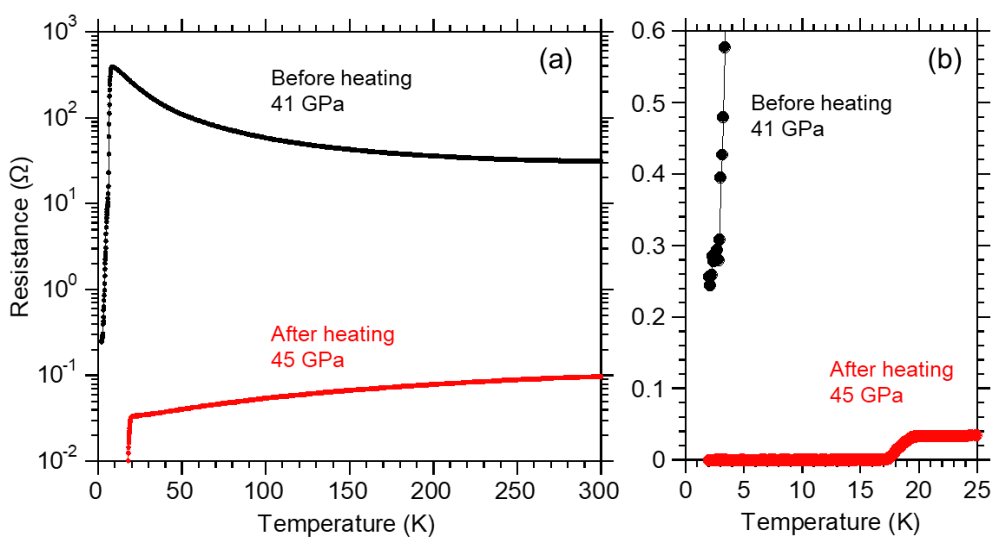

**FIG. S3.** (a) Temperature dependence of resistance in as-pressed $In_2S_3$ and obtained $In_{3-x}S_4$ under each pressure with log scale in resistance. (b) Enlarged plots around low temperature with liner scale in resistance.